# Adsorption component of the disjoining pressure in thin liquid films

Roumen Tsekov
Department of Physical Chemistry, University of Sofia, 1164 Sofia, Bulgaria

The disjoining pressure isotherm in foam films is theoretically studied and an important contribution of adsorption is discovered. On the basis of the interfacial thermodynamics an adsorption disjoining pressure component is derived, which is repulsive and exponentially decaying by the film thickness. Expressions for its magnitude and decay length are derived in terms of well-known thermodynamic characteristics such as the partial Gibbs elasticity and adsorption length. Several adsorption isotherms are considered and the corresponding adsorption disjoining pressure components are calculated. An important coupling between the interfacial layer phases and wetting transitions is discussed as well.

Thin liquid films are anisotropic structures and the pressure tensor inside a film possesses distinct components [1]. The normal component $P_N$ of the pressure tensor is constant equal to the pressure in the neighboring homogeneous bulk phases. The tangential component $P_T$ of the pressure tensor exhibits non-monotonous profile across the film. If the liquid film is thin enough, the transition regions of the two film surfaces overlap thus leading to appearance of disjoining pressure [2]. The overall excess of $P_N$ against $P_T$ defines the film tension $\gamma$ [3]

$$\gamma = \int_{-\infty}^{\infty} (P_N - P_T) dz = 2\sigma + \Pi h \qquad (1)$$

where the interfacial part of $\gamma$ equals to twice the film surface tension $\sigma$ and $\Pi$ is the disjoining pressure (see Fig. 1).

The Gibbs-Duhem relation for the film at constant temperature reads [4]

$$2d\sigma = -\Pi dh - 2\sum_i \Gamma_i d\tilde{\mu}_i \qquad (2)$$

where $\{\Gamma_i, \tilde{\mu}_i\}$ are the adsorptions and electrochemical potentials of all the film components excluding the solvent, whose adsorption is set to be zero. Equation (2) provides straightforward an important definition of the disjoining pressure as the thickness derivative of the film surface tension

$$\Pi = -2(\partial_h \sigma)_{T, \tilde{\mu}} \qquad (3)$$

as well as the following Maxwell relationship

$$2(\partial_h \Gamma_i)_{T,\tilde{\mu}} = (\partial_{\tilde{\mu}_i} \Pi)_{T,h} \tag{4}$$

which hints already the important contribution of adsorption on the disjoining pressure [5, 6]. In this paper a shorter notation $\partial_x \equiv \partial/\partial x$, being used often in the physics literature, is adopted everywhere for partial derivatives.

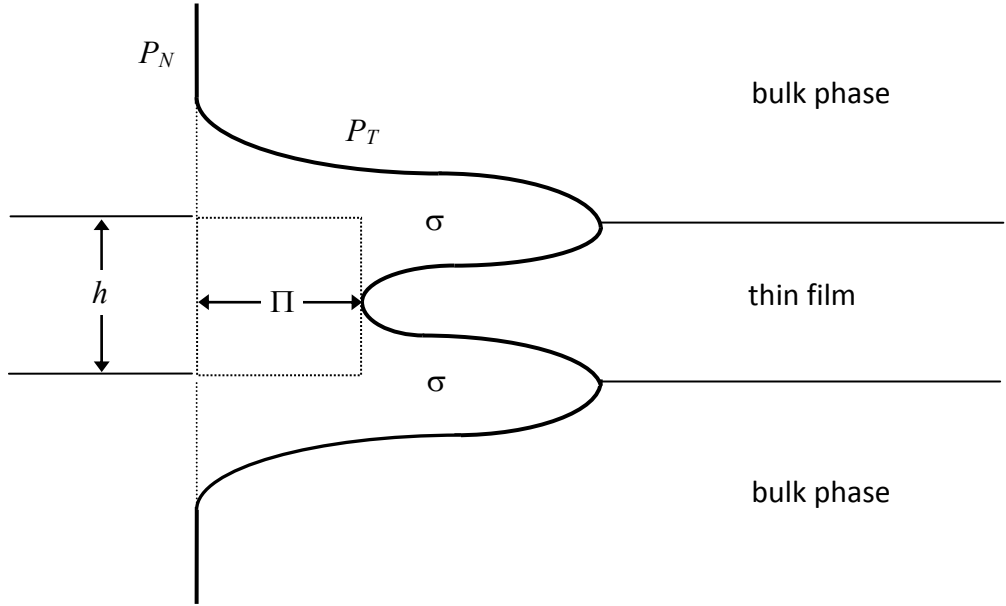

**Fig. 1** Schema of the pressure tensor distribution across a symmetric thin liquid film.

### Disjoining Pressure Isotherm

According to Eq. (2) the film surface tension depends on temperature, film thickness and electrochemical potentials of dissolved species. In theory, however, it is more suitable to consider $T$, $h$ and adsorptions as variables, i.e. $\sigma = \sigma(T,h,\Gamma)$. Hence, the film surface tension depends on the film thickness either explicitly or via the thickness dependence of the adsorptions. Thus, the definition (3) can be split into two distinct components of the disjoining pressure

$$\Pi = -2(\partial_h \sigma)_{T,\Gamma} - 2\sum (\partial_{\Gamma_i} \sigma)_{T,h} (\partial_h \Gamma_i)_{T,\tilde{\mu}} \tag{5}$$

Considering small deviations of the adsorption on the film surfaces with respect to the case of a single interface, one can replace by $(\partial_{\Gamma_i} \sigma)_{T,h=\infty} = -E_{G,i}/\Gamma_i(h=\infty) - z_i F\phi/2$ the term $(\partial_{\Gamma_i} \sigma)_{T,h}$ in Eq. (5), where $E_{G,i}$ and $\phi$ are the partial Gibbs elasticity and the surface potential on a single surface, respectively. Thus, Eq. (5) acquires the approximate form

$$\Pi = -2(\partial_h \sigma)_{T,\Gamma} + \phi(\partial_h q)_{T,\tilde{\mu}} + 2\sum E_{G,i}(\partial_h \Gamma_i)_{T,\tilde{\mu}} / \Gamma_i(h=\infty) \tag{6}$$

where $q \equiv \sum z_i F \Gamma_i$ is the surface charge density.

Obviously the first term in Eq. (6) accounts for the solvent interactions. In the case of foam film from pure solvent $-2(\partial_h \sigma)_{T,\Gamma=0}$ reduces to the van der Waals component of the disjoining pressure $\Pi_{VW} = -A/6\pi h^3$ with $A$ being the corresponding Hamaker constant [7]. Since the films consist usually by dilute solutions, their Hamaker constants do not differ substantially for those of pure solvents. One can easily recognize the electrostatic disjoining pressure component in the second term of Eq. (6). Indeed, substituting the well-known expression for the surface charge density, $q = \varepsilon_0 \varepsilon \kappa \phi \tanh(\kappa h/2)$, following from the linearized Poisson-Boltzmann equation, yields the classical expressions for the electrostatic disjoining pressure at low surface potentials or surface charge densities, respectively, [8]

$$\Pi_{EL} = \frac{\varepsilon_0 \varepsilon \kappa^2 \phi^2}{2\cosh^2(\kappa h/2)} = \frac{q^2}{2\varepsilon_0 \varepsilon \sinh^2(\kappa h/2)} \tag{7}$$

where $\kappa$ is the reciprocal value of the Debye length.

Let us consider now the last adsorption term of the disjoining pressure in Eq. (6). To calculate its thickness dependence one can employ the Maxwell relation (4). Introducing the following definition for the difference $\Delta X \equiv X(h) - X(h=\infty)$ between the values of a property in the film and in the bulk liquid one can write that $\Pi = -\Delta p$ where $p$ is the film pressure. Thus, the Maxwell relation (4) can be consecutively modified to

$$2(\partial_h \Gamma_i)_{T,\tilde{\mu}} = (\partial_{\tilde{\mu}_i} \Pi)_{T,h} = -\Delta(\partial_{\tilde{\mu}_i} p)_{T,h} = -\Delta C_i = -\alpha_i \Delta \Gamma_i \tag{8}$$

where $\{C_i\}$ are the concentrations of components and $\{\alpha_i\}$ are the reciprocal values of adsorption lengths depending on the adsorption equilibrium constants. Solving the linear differential equation (8) under the obvious condition $\Gamma_i(h=0) = 0$ yields

$$\Gamma_i = \Gamma_i(h=\infty)[1-\exp(-\alpha_i h/2)] \tag{9}$$

As is seen, the adsorption in films is lower than the adsorption on a single interface due to overlap of the concentration transition regions. Now substituting this expression into the definition of the adsorption disjoining pressure, i.e. the last term of Eq. (6), leads to

$$\Pi_{AD} = \sum \alpha_i E_{G,i} \exp(-\alpha_i h/2) \tag{10}$$

The similarity of this expression with the electrostatic Eq. (7) at large $\kappa h$ is due to the existence of adsorption diffusive layers at the film surfaces [9].

## Adsorption Disjoining Pressure

To describe the contribution of the specifically adsorbed species knowledge for the corresponding adsorption isotherms is required. Henry's isotherm $K_i C_i = \Gamma_i$ is the simplest model, which is valid for dilute solutions. In this case $\alpha_i = 1/K_i$ is a constant and the partial Gibbs elasticity depends linearly on the concentration, $E_{G,i} = RTK_i c_i$, where $\{c_i \equiv C_i(h=\infty)\}$ are the surfactant concentrations of the solution used to make the films. Introducing these expressions in Eq. (10) yields the adsorption disjoining pressure component in the form

$$\Pi_{AD} = RT \sum c_i \exp(-h/2K_i) \tag{11}$$

As is seen, $\Pi_{AD}$ is positive and represents the osmotic pressure difference between the meniscus liquid and film, which is due to different concentrations equilibrated by different adsorptions. A typical value for the Henry constant $K_i$ is of the order of hundreds of microns. This means that the adsorption disjoining pressure from Eq. (11) is extremely long-ranged repulsion. However, since the Henry isotherm is valid for very dilute solutions, the concentrations $\{c_i\}$ should be very small and thus $\Pi_{AD}$ from Eq. (11) is practically negligible.

As is expected the adsorption disjoining pressure becomes important for relatively concentrated solutions. In this case, however, the Henry isotherm is no more applicable. A reasonable model here is the Langmuir adsorption isotherm, which for the case of presence of a single surfactant reads

$$KC = \Gamma/(1-a\Gamma) \tag{12}$$

where $a$ is the molar area in the saturated layer. The partial Gibbs elasticity of the Langmuir layer is the same as above, $E_G = RTKc$, but the adsorption length $1/\alpha = K/(1+aKc)^2$ is substantially decreased. For instance, at concentrations about 1 mM the adsorption length is of the order of hundreds of Angstroms. This value is commensurable with the Debye length at the same electrolyte concentrations. Also the pre-exponential factor is significantly increased with respect to Eq. (11). Hence, at large surfactant concentrations the adsorption disjoining pressure

$$\Pi_{AD} = RTc(1+aKc)^2 \exp[-(1+aKc)^2 h/2K] \tag{13}$$

becomes important. Note that $\Pi_{AD}$ exhibits now a maximum with respect to the surfactant concentration. Equation (13) reduces naturally to Eq. (11) at low concentrations.

Since in the example above the meaningful adsorption corresponds to about 90 % of the maximal one, additional complications can arise from the interactions between the adsorbed molecules. They could be accounted for by a more complicate adsorption isotherm, such as the Frumkin one

$$KC = \exp(-\beta a\Gamma)\Gamma/(1-a\Gamma) \tag{14}$$

where the coefficient $\beta$ accounts for the interactions in the adsorption layer. The corresponding partial Gibbs elasticity and reciprocal adsorption length can be expressed as functions of the adsorption

$$E_G = RT\Gamma\frac{1-\beta a\Gamma(1-a\Gamma)}{1-a\Gamma} \qquad \alpha = \exp(-\beta a\Gamma)\frac{1-\beta a\Gamma(1-a\Gamma)}{K(1-a\Gamma)^2} \tag{15}$$

and replaced in the disjoining pressure model (10) the latter acquires the form

$$\Pi_{AD} = RT\frac{\Gamma[1-\beta a\Gamma(1-a\Gamma)]^2}{K(1-a\Gamma)^3}\exp\{-\beta a\Gamma - \exp(-\beta a\Gamma)\frac{[1-\beta a\Gamma(1-a\Gamma)]}{2K(1-a\Gamma)^2}h\} \tag{16}$$

The relevant concentration dependence can be obtained by combining this expression with Eq. (14). In the case of a positive and sufficiently large $\beta$ the Frumkin isotherm describes a phase-transition in the interfacial layer which reflects in an abrupt change of the adsorption. Thus in the two-phase coexistence region $\Pi_{AD}$ will possess two different values, which will induce film stratification. Hence, the interfacial and wetting phase-transitions are coupled and cross-linked effects are expected to occur [10].

Finally, the adsorption disjoining pressure is very sensitive to the interfacial properties. Any surface modifications will reflect in changes of the adsorption and, hence, in $\Pi_{AD}$. This is especially significant for wetting films, where the so-called hydrophobic disjoining pressure $\Pi_{HP}$ is observed [11]. It decays exponentially with the film thickness increase and has short- and long-ranged components [12]. A previous analysis [13] showed that the adsorption disjoining pressure is an important part of $\Pi_{HP}$. In contrast to the present case, however, $\Pi_{AD}$ in wetting films could be either attractive or repulsive depending on the hydrophobicity of the two film surfaces. A more detailed description of the adsorption component of the disjoining pressure could take into account the dependence of the specific adsorption constants on the film thickness as well, which is week, in general [6].